# Observation of Extraordinary Vibration Scatterings Induced by Strong Anharmonicity in Lead-Free Halide Double Perovskites

Guang Wang[1#], Jiongzhi Zheng[2#], Jie Xue[3#], Yixin Xu[1], Qiye Zheng[1], Geoffroy Hautier[2], Haipeng Lu[3] and Yanguang Zhou[1*]

[1]*Department of Mechanical and Aerospace Engineering, The Hong Kong University of Science and Technology, Clear Water Bay, Kowloon, Hong Kong SAR*

[2]*Thayer School of Engineering, Dartmouth College, Hanover, New Hampshire 03755, USA*

[3]*Department of Chemistry, The Hong Kong University of Science and Technology, Clear Water Bay, Kowloon, Hong Kong SAR*

**Abstract**

Lead-free halide double perovskites provide a promising solution for the long-standing issues of lead-containing halide perovskites, i.e., the toxicity of Pb and the low stability under ambient conditions and high-intensity illumination. Their light-to-electricity or thermal-to-electricity conversion is strongly determined by the dynamics of the corresponding lattice vibrations. Here, we present the measurement of lattice dynamics in a prototypical lead-free halide double perovskite, i.e., $Cs_2NaInCl_6$. Our quantitative measurements and first-principles calculations show that the scatterings among lattice vibrations at room temperature are at the timescale of ~ 1 ps, which stems from the extraordinarily strong anharmonicity in $Cs_2NaInCl_6$. We further quantitatively characterize the degree of anharmonicity of all the ions in the single $Cs_2NaInCl_6$ crystal, and demonstrate that this strong anharmonicity is synergistically contributed by the bond hierarchy, the tilting of the $NaCl_6$ and $InCl_6$ octahedral units, and the rattling of $Cs^+$ ions. Consequently, the crystalline $Cs_2NaInCl_6$ possesses an ultralow thermal conductivity of ~0.43 W/mK at room temperature, and a weak temperature dependence of $T^{-0.41}$. Our findings here uncovered the underlying mechanisms behind the dynamics of lattice vibrations in double perovskites, which could largely benefit the design of optoelectronics and thermoelectrics based on halide double perovskites.

---

[#] These authors contribute equally. [*]Author to whom all correspondence should be addressed. Email: maeygzhou@ust.hk

## INTRODUCTION

Hybrid and all-inorganic lead halide perovskites have been widely studied in optoelectronics (*1–4*), photovoltaics (*5–8*), and thermoelectric applications (*9*) due to their high light-to-electricity (*10, 11*), ultra-low thermal conductivity (*12–14*), high charge mobility (*15, 16*) and high Seebeck coefficients (*17*). They possess a general formula of $ABX_3$, where A denotes an organic cation such as methylammonium ($MA^+$) or formamidinium ($FA^+$), or $Cs^+$ ion, B is divalent $Pb^{2+}$ or $Sn^{2+}$, and X is the halide anion such as $Cl^-$, $Br^-$, and $I^-$. Compared to conventional semiconductors such as crystalline silicon, the crystal structure of the lead halide perovskites is special. It has a 3D-corner-sharing $[PbX_6]^{4-}$ octahedra framework, forming a cage occupied by free $A^+$ cations. This unique cage-like crystalline configuration results in nontrivial electrical, thermal, and chemical properties in lead halide perovskites. which makes them superior for various applications mentioned above. However, disadvantages such as toxicity, sensitive to oxygen, water moisture and high temperature, strongly affect their long-term stability and wider adoption (*18*).

The recently proposed lead-free halide double perovskites (HDPs) $A_2B^+B^{3+}X_6$, where the B sites are replaced by pairs of monovalent and trivalent metal cations such as $Ag^+$, $Na^+$, $Bi^{3+}$, and $In^{3+}$, are promising to resolve these issues. As this structure offers more material diversity by combining different $A^+$, $B^+$, $B^{3+}$, and $X^-$ ions, many double perovskites are proven to be non-toxic and highly stable against temperature rise and humidity (*18–21*). In addition, the high carrier mobilities and large Seebeck coefficient with low thermal conductivity make them attractive for thermoelectric applications(*22, 23*). For instance, $Cs_2AgBiBr_6$, which is one of the most intriguing HDPs with a long carrier lifetime (*24*) and small carrier effective mass (*25*) shows great potential for practical long-term applications in solar cells and X-ray detectors (*18, 26*). However, the large bandgap (i.e., > 2 eV) and strong electron-phonon coupling that affects the charge-carrier mobilities in HDPs limits their real applications in optoelectronics (*27–30*).

These challenges are closely related to the complex lattice dynamics and phonon scattering in HDPs. HDPs are generally regarded as highly anharmonic materials due to their complicated lattice structure. Their low thermal conductivity (e.g., ~0.21 W/mK at room temperature for $Cs_2AgBiBr_6$ (*31*)), although beneficial for thermoelectrics, will inevitably result in a high-temperature rise in the device during operation which affects the quantum efficiency and lifetime. Therefore, understanding the dynamics of lattice vibrations, which guides the regulation of the band structure and thermal transport properties of HDPs, is critical for designing high-performance HDPs-based devices.

Microscopically, it is known that the $A^+$ cations in perovskites can diffuse largely in cages formed by octahedra while other ions vibrate with a small amplitude around their equilibrium sites. The unique ordered-disordered character of HDPs with ordered sublattices and A-site cations exhibiting confined disordered diffusion yields exceptional lattice dynamics in HDPs. For example, first-principles calculations show that the tendency toward octahedral tilting instability of $Cs_2AgBiBr_6$ is caused by the strong anharmonicity (*31*, *32*), differing from the behavior in single perovskites where the tilting stems from the undersize $A^+$ cations that can be described by the Goldschmidt tolerance factor (*33*). However, detailed studies of the dynamics of lattice vibrations in HDPs remain limited. A deeper understanding is highly desirable to chart how phonon anharmonicity benefits thermal transport and whether the vibrations of the sublattice are modulated by $A^+$ cations. A major obstacle to clarifying the lattice dynamics of HDPs lies in the challenges of synthesizing crystals with large grain sizes and the high computational costs required for first-principles simulations. These challenges have impeded our understanding of the ultralow lattice thermal conductivity and the establishment of a clear physical picture of the coupling of sublattice vibrations and A site cations' anharmonicity in HDPs.

Herein, we thoroughly investigate the nature of the lattice vibrations in a novel lead-free HDP, $Cs_2NaInCl_6$, which possesses a direct band gap (~2.73eV) (*34–36*) as opposed to the indirect bandgap in $Cs_2AgBiBr_6$. $Cs_2NaInCl_6$ has a similar structure as $Cs_2AgBiBr_6$, and therefore, exhibits a strong signature of scattering and phonon coherence effect caused by the strong anharmonicity. By combining experimental thermal transport and Raman study with first-principles calculations, the corresponding lattice thermal conductivity is experimentally measured to be as low as ~0.43 W/mK at room temperature due to the strong scatterings which also shows a weak temperature dependence of $T^{-0.41}$ owing to the large contribution from the coherence channel. Our first-principles calculations further demonstrate that this strong anharmonicity in HDPs stems from the synergetic mechanism between the tilting of $InCl_6$ and $NaCl_6$ octahedra and the rattling vibration of $Cs^+$ ions.

## RESULTS AND DISCUSSIONS

**Materials synthesis and characterization**. The structure of $Cs_2NaInCl_6$ is similar to lead-based halide perovskite and possesses a cubic structure at room temperature. It can be regarded as the $Pb^{2+}$ sites being replaced by $Na^+$ and $In^{3+}$ ions in an ordered manner (**Figure 1a** and **b**). Here, the high-quality $Cs_2NaInCl_6$ crystals with grain sizes from hundreds of micrometers to millimeters are synthesized by the solution cooling method (*34*) (see **Methods** for details). Room temperature powder X-ray diffraction (PXRD) was conducted to confirm the purity and phase of the as-synthesized HDPs, which agrees well with the calculated results using the crystallographic information file (*34*) (**Figure 1c**). The lattice constant of the as-synthesized HDPs obtained from the refinements is ~10.53 ± 0.005 Å ($R_{wp}$=13.05%) and agrees well with the reference (*37*). The scanning electron microscopy (SEM) images show a good regular crystal morphology and large grain size of the $Cs_2NaInCl_6$ (**Figure 1c**), which can benefit the later frequency domain thermoreflectance (FDTR) measurements of thermal conductivity. The distinct lattice patterns and regular diffraction spots obtained using high-resolution transmission electron microscopy (HRTEM) measurements further indicate the defect-free single crystallinity of synthesized $Cs_2NaInCl_6$ (**Figures 1d** and **1e**). To reveal the lattice dynamics, we use Raman spectroscopy to gain insight into the optical phonon modes and the temperature-dependent anharmonicity, as shown in **Figure 1f**. The peaks at 141.3 and 294.6 cm$^{-1}$ can be attributed to the bending ($T_{2g}$) and stretching ($A_{1g}$) of $[InCl_6]^{3-}$ octahedron, respectively (*38*). As there is no significant difference in the spectrums under different excitation laser wavelengths (785 nm, 633 nm, 514 nm), the 633 nm laser is used in the temperature-dependent Raman measurements hereafter.

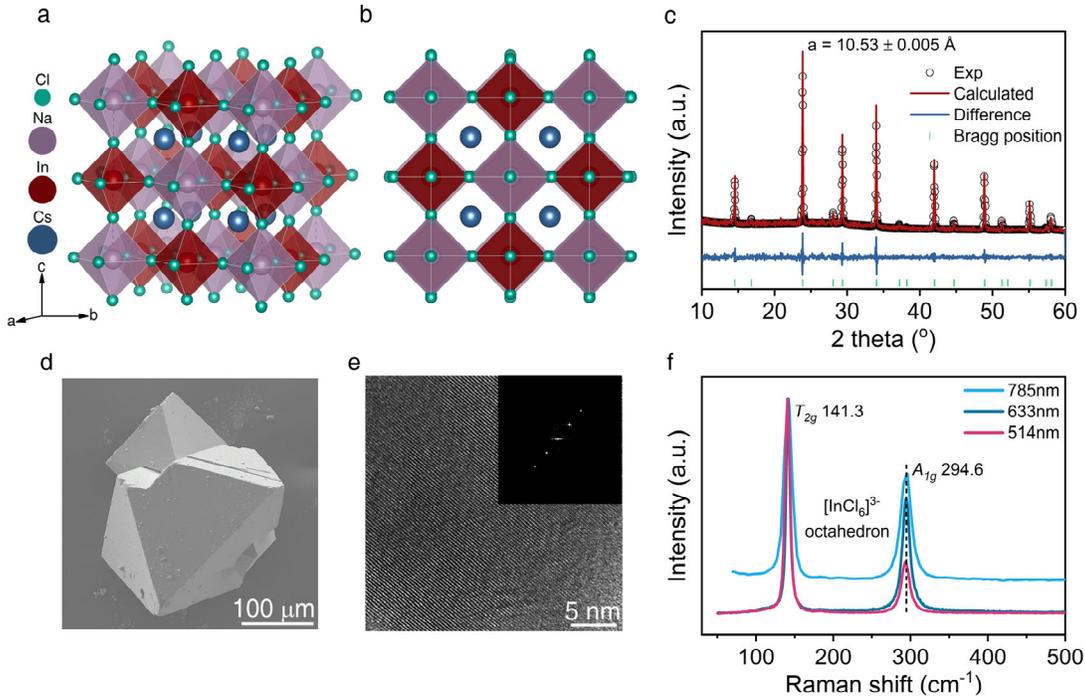

**Figure 1 Characterization of $Cs_2NaInCl_6$.** a, b. The schematic of the $Cs_2NaInCl_6$ structure drawn by Vesta (*39*). c. Representative refined PXRD results of $Cs_2NaInCl_6$ powder at room temperature. The difference between the raw data (black circles) and the calculation (red curve) is the dark blue curve. d. The SEM image of $Cs_2NaInCl_6$ single crystal with a size of ~200 $\mu$m. e. The high-resolution TEM images of $Cs_2NaInCl_6$ crystals and the corresponding fast Fourier transform (FFT) diffraction spots of the lattice pattern as shown in d (inset figure). f. The Raman spectrum of $Cs_2NaInCl_6$ single-crystal excited at different laser wavelengths.

**Thermal conductivity**. The thermal conductivity of $Cs_2NaInCl_6$ crystals was measured by the well-established optical pump-probe FDTR method (*40, 41*). These crystals with smooth and flat surfaces were selected under a 10× optical microscope to ensure good thermoreflectance signals (see **Methods** for details). A gold layer with a thickness of ~100 nm (**Figure S1**) was deposited on the samples as the transducer, which absorbs the modulated pump laser and enables the monitoring of the corresponding rapid surface temperature oscillation by thermoreflectance effect at the probe laser wavelength which was detected by a lock-in amplifier. The phase lag between the pump and probe lasers was then analyzed and fitted to a bi-layer heat diffusion model to obtain the thermal conductivity of the sample (*42, 43*). The $1/e^2$ radii of the pump and probe lasers in our system are ~22 $\mu$m and ~5 $\mu$m, respectively, which was measured by a beam offset method each time before the FDTR measurement. The samples were placed on a compact temperature control stage equipped with a semi-sealed chamber with gas purging capability. The environment temperature of the stage can be tuned continuously from 873K to 77 K using a liquid nitrogen circuit. It is noted that the thermal properties of the Au transducer deposited on the samples are temperature dependent which can influence the measurement. We therefore first measured the temperature-dependent thermal conductivity and heat capacity of the Au transducer deposited on a fused silica by FDTR (see **Methods** for details).

The thermal conductivity of $Cs_2NaInCl_6$ from 273 K to 413 K was measured (see **Methods for details**), as shown in **Figure 2a**. It is noted that the heat capacity and thermal conductivity of $Cs_2NaInCl_6$ have similar phase sensitivity (**See SI Note 2 for details**) across the whole frequency range in the fitting, we therefore use the heat capacity obtained by DFT calculations (**Figure S4**) as the input value and only relax the thermal conductivity of the $Cs_2NaInCl_6$ in the nonlinear regression to ensure the accuracy of the data. The derived heat capacity of the material at high temperatures converges to the classical value from the Dulong-petit law (*48*). Our results show that the thermal conductivity of $Cs_2NaInCl_6$ is lower than a wide range of typical perovskites, such as $CsSnI_3$ and $MAPbI_3$, reaching ~0.43 W/mK at room temperature, and is comparable to $CsPbBr_3$ and $CsPbI_3$, as shown in **Figure 2b**. However, the measured thermal conductivities show a weaker temperature dependence of $T^{-0.41}$ compared to other perovskites and significantly deviate from the common $T^{-1}$ dependence observed in weakly anharmonic solids (*49*, *50*). It may be deduced that the ultra-low thermal conductivity of $Cs_2NaInCl_6$ stem from the strong anharmonicity as observed in other HPDs (*51*, *52*).

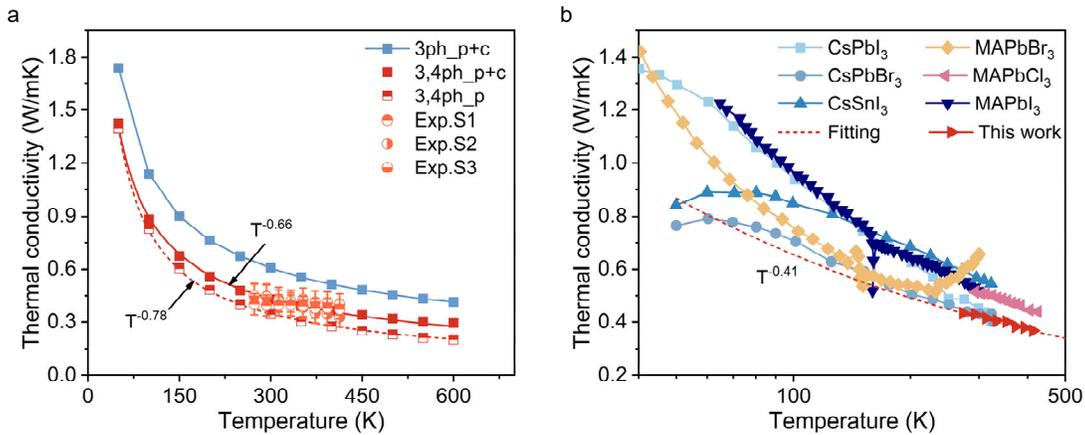

**Figure 2 The thermal conductivity of $Cs_2NaInCl_6$ from experiment and DFT calculations.** a. The temperature-dependent thermal conductivity of $Cs_2NaInCl_6$ obtained by FDTR measurements (three typical samples are denoted as Exp.S1, Exp.S2 and Exp.S3), and DFT calculations (with phonon renormalization), including both the coherence and population contributions. b. The thermal conductivity of $Cs_2NaInCl_6$ compared with other typical perovskites, including $CsPbI_3$ (*44*), $CsPbBr_3$ (*44*), $CsSnI_3$ (*44*), $MAPbBr_3$ (*45*), $MAPbCl_3$ (*46*) and $MAPbI_3$ (*47*).

To comprehensively elucidate the microscopic mechanisms of thermal transport, we also calculated the thermal conductivity of $Cs_2NaInCl_6$ using the state-of-the-art first-principles-based unified theory (*53*), in which both the population's and coherence's contributions are considered. The calculated results show that only considering three-phonon (3ph) scatterings overestimates the experimental thermal conductivity (**Figure 2a**), which indicates that the four-phonon (4ph) scatterings should also be included in our calculations. However, it is interesting to find that the calculated thermal conductivity of $Cs_2NaInCl_6$ considering only population is lower than the measured values, which highlights the importance of coherence's contribution to the total thermal conductivity. The calculated thermal conductivity including the coherence channel agrees well with the measured results, and a weak temperature dependence of $T^{-0.66}$ is also observed.

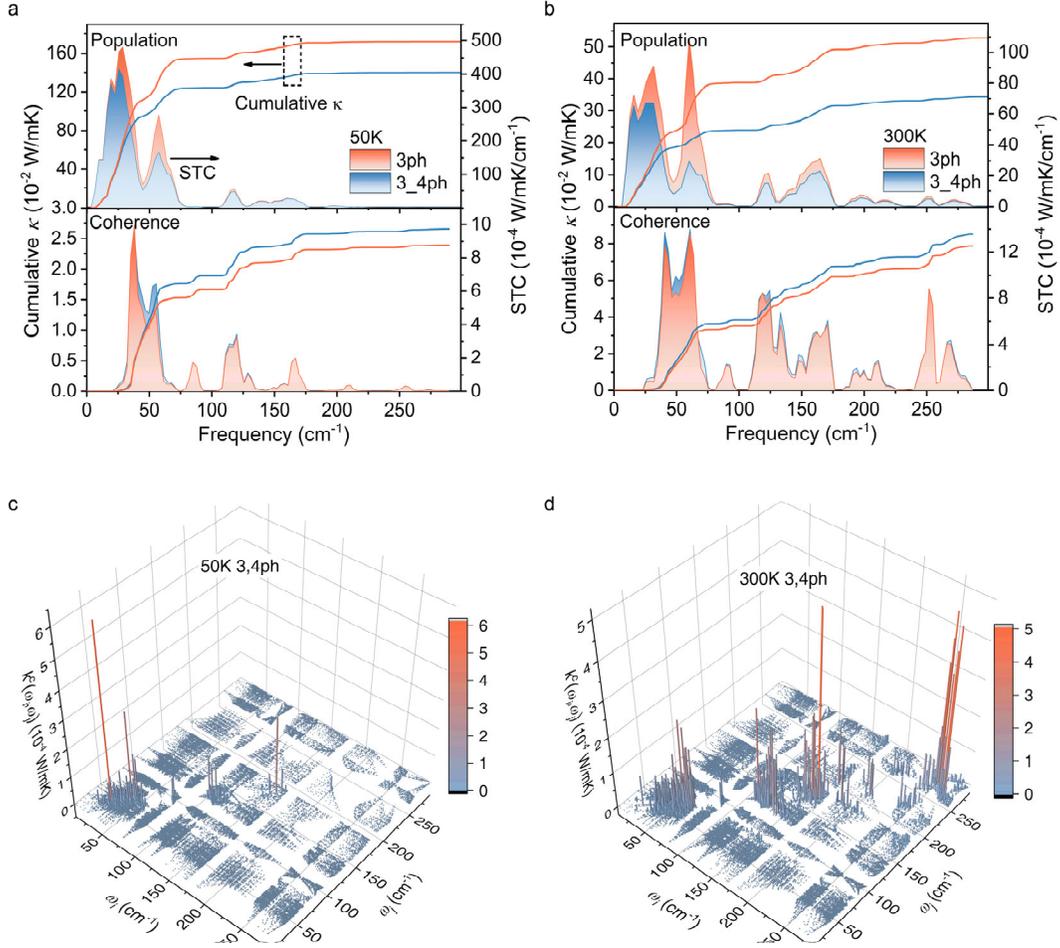

**Figure 3 Calculated population's and coherences' thermal conductivity. a.** Calculated cumulative and spectral wave-like phonon conductivity at 50 K considering only 3ph scatterings and both 3ph and 4ph scatterings, respectively. **b.** The same as (a) but at 300 K. **c.** The calculated two-dimensional modal coherences' thermal conductivity considering both 3ph and 4ph scatterings at 50 K. **d.** The same as (c) but at 300 K.

**Two-channel thermal transport**. We then further investigate the thermal transport mechanisms in crystalline $Cs_2NaInCl_6$ at 50 K and 300 K by analyzing both the phonon population's and coherence's contributions. It is known that the thermal energy will be transferred through phonon propagation (*54*, *55*). Meanwhile, due to the broaden phonon linewidth, an overlap occurs between the densely packed phonon branches, which is regarded as phonon coherence. Consequently, the thermal energy will also be transferred through the phonon coherence channel (*53*, *56–58*). At a low temperature of 50 K, the primary contribution to the propagations' conductivity arises from phonons with frequencies below ~80 cm$^{-1}$, as depicted in **Figure 3a**. Specifically, owing to the strongly intrinsic anharmonic scatterings (see detailed analysis below), the thermal conductivity of Cs$_2$NaInCl$_6$ resulting from phonon propagation is ~1.399 W/mK, while the coherence thermal conductivity contributed by the wave-like tunnelling is relatively small and has a value of ~0.027 W/mK (**Figures 3a**). A significant portion of the population's contribution can be attributed to the fact that many phonons remain unexcited at low temperatures. When the temperature increases to 300 K, all the phonons in Cs$_2$NaInCl$_6$ are excited and contribute to the thermal transport (**Figure 3b**). The thermal conductivity contributed by the phonon propagation is found to decrease to ~0.344 W/mK as the phonon scatterings become stronger. Conversely, coherence's conductivity increases from ~ 0.027 W/mK at 50 K to ~ 0.085 W/mK at 300 K, underscoring its increasingly critical role in the total thermal conductivity, particularly at higher temperatures. Correspondingly, its contribution to the total thermal conductivity rises from around 1.9% at 50 K to 20% at 300 K, and it continues to increase at higher temperatures (**Figure 3a** and **b**). More specifically, as the temperature increases, an increasing number of high-frequency phonon and phonon pairs with notably different frequencies contribute to the coherence's conductivity (**Figure 3c and d**). This observation can be attributed to the diminishing particle-like nature of phonons at higher temperatures. It is known that the thermal conductivity of a crystal contributed by phonon propagation follows a $T^{-1}$ relationship if the Umklapp 3ph scatterings are dominant in that crystal (*49*, *59*). With the inclusion of 4ph scatterings, the propagation's conductivity has a weaker temperature dependence which follows the ~$T^{-0.78}$ relationship (**Figure 2a**). However, as more phonons transport thermal energy through wave-like tunnelling, the contribution of coherence thermal conductivity becomes non-negligible. Therefore, the total thermal conductivity of $Cs_2NaInCl_6$ exhibits a $\sim T^{-0.66}$ temperature dependence, which results from the interplay between the propagation and coherence channels.

To reveal the relationship between the low thermal conductivity and lattice anharmonicity, we further calculate the degree of anharmonicity (DOA) of HDPs, which can be used to quantitatively measure the anharmonicity of materials (*60, 61*). The DOA of a material is defined as

$$\sigma^A(T) = \frac{\sigma[F^A]_T}{\sigma[F]_T} = \sqrt{\frac{\sum_{I,\alpha}\left\langle (F_{I,\alpha}^A)^2 \right\rangle_T}{\sum_{I,\alpha}\left\langle (F_{I,\alpha})^2 \right\rangle_T}} \tag{1}$$

where $I$, $\alpha$, and $T$ are atomic index, Cartesian direction, and temperature, respectively. $\sigma^A(T)$ measures the degree of lattice anharmonicity by normalizing the deviation of anharmonic atomic forces $\sigma[F^A]_T$ to total atomic forces $\sigma[F]_T$. We calculated the DOA of $Cs_2NaInCl_6$, $Cs_2AgBiBr_6$, and $CsPbI_3$ at different temperatures. For comparison, the DOA of some other well-known strong anharmonic crystals were also listed, as shown in **Figure 4a**. It can be found that the experimental thermal conductivity of the regular solids exhibits a strong negative correlation with their DOA, approximately following a power-law model, i.e. $\kappa \propto (\sigma^A)^{-4.8}$ (*60, 61*). Whereas, the DOA of these complicated solids including $Cs_2NaInCl_6$ largely deviates from the power-law model at all calculated temperatures, indicating the increased variety and complexity in these systems. This anharmonicity can be reflected by the small phonon lifetime, which are less than 10 ps at 50 K and less than 1 ps at 300 K (**Figure 4f**). We also measure the lifetime of two typical phonon modes (i.e., $A_{1g} \sim 141.3$ cm$^{-1}$ and $T_{2g} \sim 294.6$ cm$^{-1}$) using the temperature-dependent Raman measurements (**Figure 4d**), in which the phonon lifetime can be deduced by the full width at half maximum (FWHM) of Raman peaks. The measured phonon lifetimes based on $\tau_i = \frac{1}{2\pi(FWHM_i)}$ (*62–64*) are ranging 0.4~1.35 ps at temperatures between 4.4 K and 300 K, which are comparable to our calculated results and phonon lifetimes observed in other halide perovskites (*51, 64*). Both our theoretical analysis and experimental measurements reveal that there is a strong anharmonicity in Cs$_2$NaInCl$_6$, which leads to its ultra-low thermal conductivity.

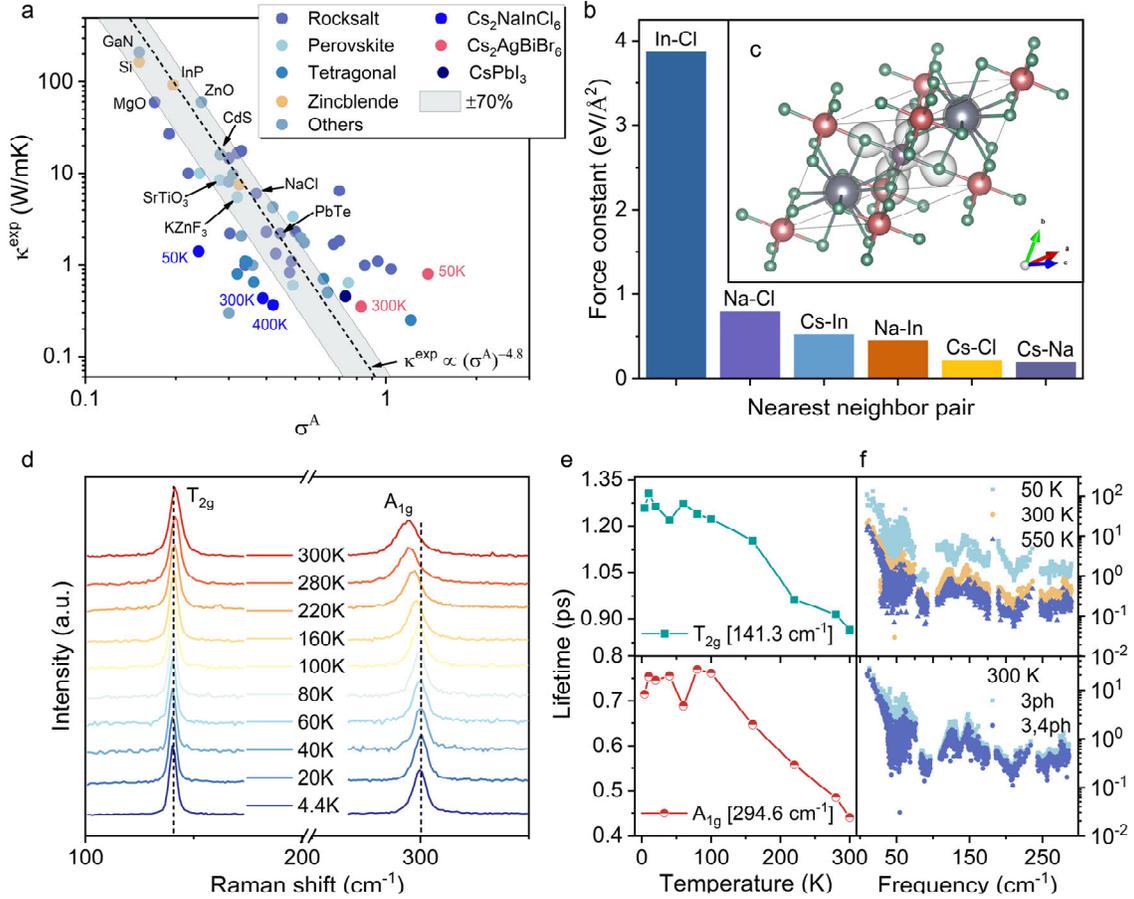

**Figure 4 The lattice vibration properties**. a. The relation between the DOA and the thermal conductivities of typical crystals at 300K. b. The harmonic interatomic force constants (IFCs) of nearest neighbor pairs in Cs$_2$NaInCl$_6$. c. The total electronic charge density in Cs$_2$NaInCl$_6$, where the iso-surfaces of charge density is denoted as a light gray color. Cs, Na, In, and Cl are shown in gray, brown, purple grey, and green spheres. d. The temperature-dependent Raman measurements of Cs$_2$NaInCl$_6$ from 4.4 K to 300 K. The peaks of T$_{2g}$ and A$_{1g}$ phonon modes are broadened with increasing temperature. e. The lifetime of T$_{2g}$ and A$_{1g}$ modes

derived from the linewidth of Raman peaks, and the DFT calculated phonon lifetimes at different temperatures considering 3ph or 3,4ph scattering process.

**The underlying mechanism for the strong anharmonicity**. Next, we reveal the underlying mechanisms responsible for the strong anharmonicity observed in both our experimental and theoretical results. Based on the atom-resolved phonon density of states (PhDOS) (**Figure 5b**) and the spectral and thermal conductivity function (**Figure 3a and b**), it can be found that the thermal transport is mainly contributed by vibrations of Cl and Cs atoms, which have large PhDOS in the low-frequency range (i.e., 0 - 100 cm$^{-1}$). These modes demonstrate strong anharmonicity, as reflected by the strong phonon stiffening in dispersion (**Figure 5a**). Specifically, the lowest-optical modes, characterized by the out-of-phase and in-phase tilting of the NaCl$_6$ and InCl$_6$ octahedral units, and rattling-like flat modes dominated by Cs ions, display a strong temperature dependence. Hence, low-frequency optical modes with strong anharmonicity will significantly suppress the acoustic modes, resulting in the low thermal conductivity in Cs$_2$NaInCl$_6$. This phenomenon was also observed in the Cs$_2$AgBiBr$_6$ crystal (*31*). The flat phonon bands associated with rattling vibrations of Cs ions usually contribute to strong 4ph scattering rates due to the large scattering phase space (*65, 66*), which is also observed in AgCrSe$_2$ (*67*) and Cu$_{12}$Sb$_4$S$_{13}$ tetrahedrites (*65*). Therefore, including 4ph scattering results in a considerable reduction in thermal conductivity (**Figure 2a**).

To gain deeper insights into the structure-related lattice anharmonicity, we calculate the harmonic interatomic force constants (IFCs) involving Cl or Cs ions (**Figure 4b**), which can be used to describe the bond strength. The IFCs show a significant difference among In-Cl, Na-Cl, Cs-Cl, Cs-Na, and Cs-In pairs. For instance, the IFC of the In-Cl pair is ~3.87 eV/Å$^2$ while is only ~0.79 eV/Å$^2$ for Na-Cl. We also analyzed the electronic charge density to understand the local chemical bonding. It is seen that the charge clouds of In and Cl are strongly overlapped, which confirms their strong covalent bonding and is formed by sharing the electrons of In and Cl atoms (**Figure 4c**). However, no charge cloud overlap is observed for Na-Cl and Cs-Cl, which indicates an absence of covalent bonding. These pairs can thus be regarded as interacting with the rest of the lattice structures by ionic bonding (*51*, *64*). The large discrepancy in the interaction strength of In-Cl and Na-Cl causes a significant bond hierarchy, and results in special lattice vibration behavior of some phonon modes containing Na and In, as Na and In atoms are at equivalent positions in HDPs. This can lead to strong anharmonicity, as verified by the atom-resolved DOA calculation. The higher DOA indicates a stronger anharmonicity (**Figure 5c**). It is found that Na and In atoms exhibit the largest DOA, which means some special phonon modes contributed mainly by Na and In atoms (e.g., the octahedral groups) have large anharmonicity. We then calculated potential energy surfaces (PES) (**Figure 5e**) of the two most typical soft modes associated with $InCl_6$ and $NaCl_6$ octahedra tilting, M1 (i.e., out-of-phase tilting) at **Γ** point and M2 (i.e., in-phase tilting) at **X** point, which are illustrated in **Figure 5d**. The PES shows a single-well potential nature which is different from the double-well potential as reported in $Cs_2AgBiBr_6$ (*31*), indicating a stable crystal structure. However, the second-order term failed to fully describe the potential energy (blue dot curve). This discrepancy reveals the anharmonic behaviours of these two modes, which require the consideration of the higher-order term's contribution (the fitting by fourth order is plotted as a green dot curve). The anharmonicity can be further quantified by the mode-resolved DOA calculated at **Γ** point, as shown in **Figure 5f**. The DOA is much larger at the low-frequency range below 100 $cm^{-1}$, which can be ascribed to the large scattering rate of corresponding vibrational modes (**Figure 4f**). For example, the anharmonic contributions of out-of-phase tilting mode (M1) make for roughly 80% of the forces, indicating its dominating effect. It is noted that the rattling-like vibration of Cs atoms (M3) also has a large anharmonic contribution, which can be attributed to the loose bonding of Cs atoms (*31*). Therefore, the tilting of the $NaCl_6$ and $InCl_6$ octahedral units, and the rattling of Cs atoms lead to significant anharmonicity in $Cs_2NaInCl_6$, which results in the ultra-low and weak temperature dependence of thermal conductivity in $Cs_2NaInCl_6$.

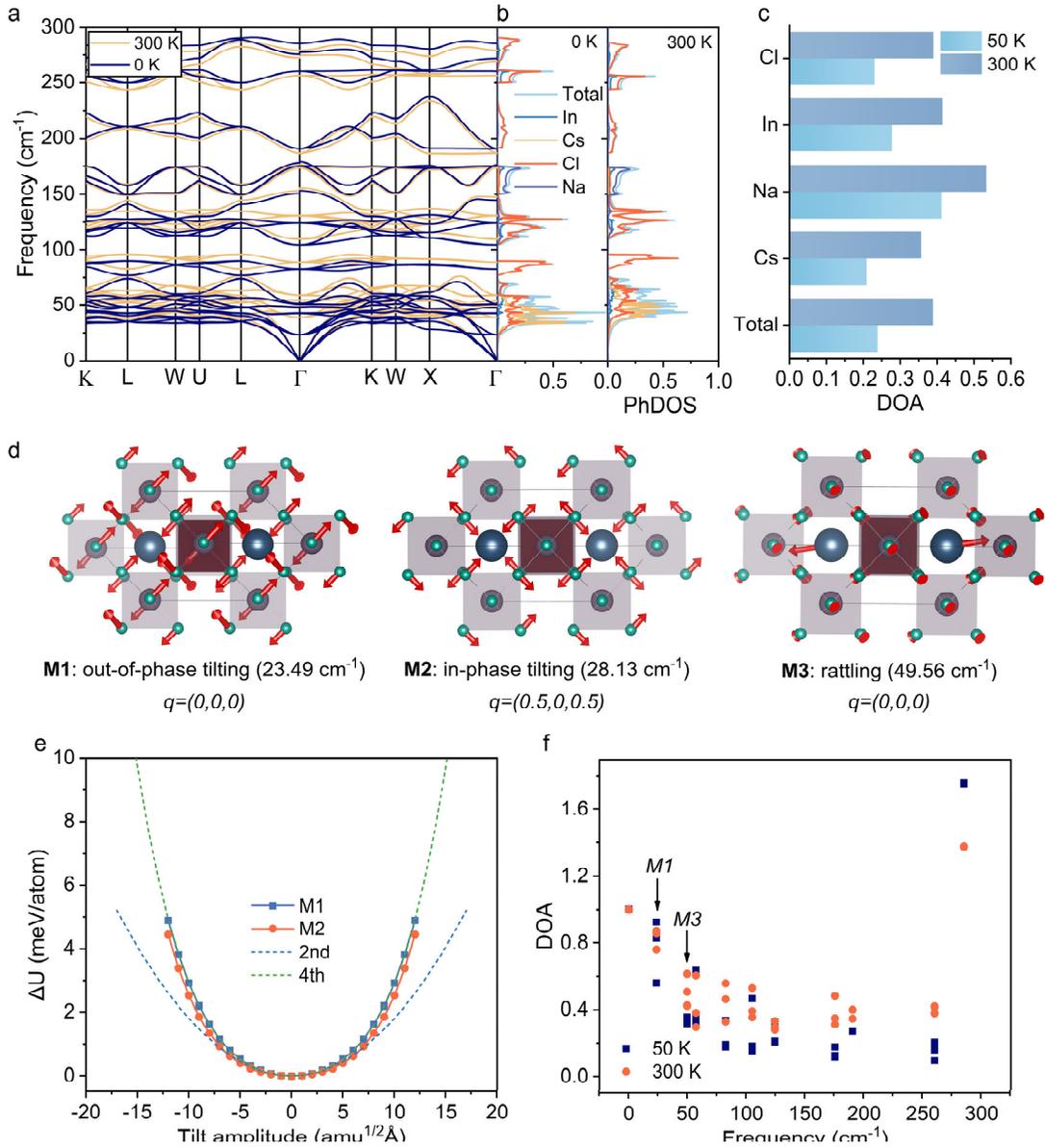

**Figure 5 The origin of vibrational anharmonicity**. a. The phonon dispersion of $Cs_2NaInCl_6$ at 0 K and 300 K. b. The atom-resolved PhDOS of $Cs_2NaInCl_6$ at 0 K and 300 K. c. The anharmonic contributions by atoms at 50 K and 300 K. d. The illustration of tilting modes for $NaCl_6$ and $InCl_6$ octahedral units and the rattling of Cs atoms. These figures are drawn by VESTA with eigenvectors adopted from Phononpy. The red arrows denote the vibration direction. It is noted that there is no vibration phase information in VESTA, and the rotation direction of $NaCl_6$ and $InCl_6$ octahedral units in in-phase tilting mode is always the same. e.

The DFT calculated potential energy surfaces of M1 and M2 modes. The blue and green dot curves indicate the potential energy surfaces decomposed to second and fourth orders, respectively. f. The mode resolved degree of anharmonicity (at $\Gamma$ point) at 50 K and 300 K.

**CONCLUSIONS**

In conclusion, we have thoroughly investigated the lattice dynamics of a prototypical lead-free halide double perovskite, i.e., $Cs_2NaInCl_6$, using both experiments and first-principles calculations. Both FDTR measurements and first-principles calculations demonstrate that the room-temperature thermal conductivity of $Cs_2NaInCl_6$ is ~0.43 W/mK. The corresponding thermal conductivity further shows a weak temperature dependence of $T^{-0.41}$ owing to the competition between the channel of the phonon population and the phonon coherence channel. Our temperature-dependent Raman measurements and anharmonic lattice dynamics calculations further find that the phonon lifetime in $Cs_2NaInCl_6$ is only ~1 ps at 300K, which implies an extraordinarily strong anharmonicity in $Cs_2NaInCl_6$. Our quantified degree of anharmonicity based on first-principles simulations shows that the extraordinarily strong anharmonicity in $Cs_2NaInCl_6$ dominantly contributed by the tilting modes of $NaCl_6$ and $InCl_6$ octahedrons and the rattling modes of Cs atoms. Our work here unveils the underlying mechanisms for the intriguing phonon dynamics and the thermal transport properties in $Cs_2NaInCl_6$ using both experimental measurements and first-principles simulations, which potentially benefits the related applications based on lead-free halide double perovskite.

## METHODS

**Synthesis and characterizations of Cs$_2$NaInCl$_6$ crystals**

Unless otherwise noted, all chemicals were used as received. Cesium chloride (CsCl, 99.9%) was purchased from Aladdin. Sodium chloride (NaCl, 99.99%) was purchased from Alfa Aesar. Indium oxide (In$_2$O$_3$, 99.99%) was purchased from Macklin. Hydrochloric acid (HCl, 37%) was purchased from Sigma-Aldrich. The solution cooling method was used to synthesize Cs$_2$NaInCl$_6$ single crystal. Firstly, 3 mmol of CsCl, 1.5 mmol of NaCl, 0.75 mmol of In$_2$O$_3$, and 10 mL of HCl were mixed, stirred, and heated to 100 °C. After heating for 1 hour, a clear solution was obtained. Then, the heating was stopped, and the solution was cooled down to room temperature at 2 °C/h to obtain the single crystals.

The powder XRD data for the finely ground Cs$_2$NaInCl$_6$ powder was collected at room temperature by powder diffractometer X'pert Pro (PANalytical, CuKα1 radiation, λ = 1.54056 Å). The morphology of the Cs$_2$NaInCl$_6$ crystal was characterized by SEM (JSM-7100F Jeol). The crystals were grounded in an agate mortar, dispersed in Ether by ultrasonication, and dropped on a copper TEM grid to obtain the high-resolution lattice information. Some thin pieces can thus be found and observed using scanning transmission electron microscopy (STEM, JEM-ARM200F JEOL). The FFT patterns of corresponding lattice fringes were obtained in DigitalMicrograph software suite.

**Temperature-dependent Raman measurements**

The temperature-dependent Raman was conducted by a Raman spectrometer (InVia, Renishaw) with a wavelength of 633 nm. Before the measurement, the Cs$_2$NaInCl$_6$ crystals were mounted on a heating stage whose temperature could be precisely controlled from liquid helium (~ 4 K) to hundreds of degrees Celsius in a vacuum chamber. There is a quartz window on the chamber to allow laser irradiation. We pumped the chamber to a low pressure overnight to avoid moisture condensation at low temperatures. During measurements, the temperature was controlled by a temperature controller with high precision. The backscattered signal was collected through 50× objective.

**FDTR measurements**

FDTR is a well-established pump-probe method to measure the thermal properties of materials. Two continuous-wave (cw) lasers were used in our apparatus. The pump laser is a 365mW (nominapower) diode laser with a wavelength of 445nm (OBIS 445-365C) which can be modulated by the lock-in amplifier (HF2LI, Zurich). A ~100 nm Au layer is chosen as the transducer to obtain a good signal-noise reflectance. The temperature rise on the Au transducer caused by the irradiation of pump laser can be detected by the probe laser with a wavelength of 532 nm (20 mW, OBIS LS532-20). The phase lag between the pump and probe lasers are analyzed by the lock-in amplifier and fitted to a heat diffusion model to obtain the interested thermal properties. We selected the sample with flat and smooth surfaces for further measurements. The Gaussian profiles of the pump laser obtained by the beam-offset method were used to check the surface quality. Only these samples with good laser profiles were chosen (*68*). Typically, the radii of pump and probe lasers were 22 $\mu$m and 5 $\mu$m, respectively.

For the temperature-dependent thermal conductivity measurement, we sputtered a gold layer on a commercial fused silica substrate at the same time with the sample. The thickness of Au film deposited was measured by atomic force microscopy, and the temperature-dependent thermal conductivity and heat capacity were obtained by FDTR test on the Au/fused silica sample. The heat capacity of $Cs_2NaInCl_6$ used in FDTR fitting was obtained by first-principles calculations. The sample was mounted on a thermal plate, which is connected to liquid nitrogen flow to reduce the temperature, in a semi-sealed chamber (Instec). For the measurement at all temperatures, the chamber was filled with dry nitrogen gas to avoid moisture condensation at sample surface. The raw phase lag curves were fitted by least mean square method to a heat diffusion model, of which the fitting error originated from the default uncertainty of ~5% of all the known parameters.

**First-principles calculations and force constants extraction**

All *ab-initio* calculations were conducted within the framework of density functional theory (DFT) (*69*), as implemented in the Vienna *Ab-initio* Simulation Package (VASP) (*70*). The projector-augmented wave (PAW) method (*71*) was employed to treat the valence states of Cs, Na, In, and Cl atoms in $Cs_2InNaCl_6$ with electron configuration of $9(5s^25p^66s^1)$, $7(2P^63s^1)$, $13(4d^{10}5s^25p^1)$ and $7(3s^23p^5)$, respectively. For both structural optimization and static calculations in $Cs_2InNaCl_6$, we utilized the revised Perdew-Burke-Ernzerhof exchange-correlation functional for solids (PBEsol) (*72*) of the generalized gradient approximation (GGA) (*73*) with a plane-wave energy cutoff of 600 eV. For structural relaxation of the unit cell, a 8×8×8 $\Gamma-center$ *k*-mesh was used to sample the Brillouin zone and the convergence criteria were set to $10^{-5}$ eV·Å$^{-1}$ for atomic forces and $10^{-8}$ eV for total energy, respectively. The fully optimized lattice constants, $a = b = c = 10.50$ Å, align well with the experimental value of *a*=10.53 Å for $Cs_2InNaCl_6$ crystal in the $Fm\overline{3}m$ space group. To account for the splitting of longitudinal (LO) and transverse optical (TO) phonon modes, and utilizing density functional perturbation theory(*74*), the calculated dielectric tensor $\varepsilon$ and Born effective charges $Z$ are determined as follows: $\varepsilon^\infty = 2.964$, $Z^*(Cs) = 1.341$, $Z^*(In) = 3.017$, $Z^*(Na) = 1.313$, $Z^*(Cl)_\perp = -0.881$, and $Z^*(Cl)_\parallel = -1.744$.

The 0-K harmonic interatomic force constants (IFCs) were extracted using the finite-displacement approach (*75*), as implemented in the **ALAMODE** package (*76*). Accurate forces were obtained from calculations performed on a 2×2×2 supercell and 4×4×4 $\Gamma-center$ *k*-mesh in VASP. To efficiently extract anharmonic IFCs, we utilized the compressive sensing lattice dynamics method (CSLD) (*77*). This technique was applied to selectively capture the most physically significant terms of the anharmonic IFCs using a limited set of displacement-force data. We initially generated 200 atomic configurations using the random-seed method by imposing a random-direction displacement of 0.15 Å on all atoms within the equilibrium 2×2×2 supercell. Utilizing the displacement-force datasets obtained from precise DFT calculations along with 0-K harmonic IFCs, we employed the least absolute shrinkage and selection operator (LASSO) technique (*78*) to extract anharmonic IFCs up to the sixth order. The real-space cutoff radii for IFCs extraction were set to 8.47 Å for cubic, 7.40 Å for quartic, 4.23 Å for quintic, and 3.18 Å for sextic interactions, respectively.

**Self-consistent phonon calculations**

To calculate the anharmonic phonon energy at finite temperatures, we employed the self-consistent phonon (SCP) calculation method, as implemented in the **ALAMODE** package (*79*). Using the quartic anharmonicity, represented by fourth-order force constants, the positive phonon energy shifts were evaluated using the following self-consistent phonon (SCP) equation:

$$\Omega_q^2 = \omega_q^2 + 2\Omega_q I_q, \tag{1}$$

where, $\omega_q$ represents the bare harmonic phonon frequency associated with phonon mode $q$, and $\Omega_q$ is the anharmonically renormalized phonon frequency at finite temperatures. The quantity $I_q$, which quantifies the influence of anharmonicity on the phonon modes, can be defined as follows:

$$I_q = \frac{1}{8N}\sum_{q'}\frac{\hbar V^{(4)}(q;-q;q';-q')}{4\Omega_q \Omega_{q'}}[1 + 2n(\Omega_{q'})], \tag{2}$$

Here, $N$ denotes the total number of sampled phonon wavevectors in the first Brillouin zone, $\hbar$ represents the reduced Planck constant, $n$ refers to the Bose-Einstein distribution, and $V^{(4)}(q;-q;q';-q')$ is the reciprocal representation of fourth-order interatomic force constants (IFCs).

To precisely evaluate phonon energy shifts at finite temperatures, inclusion of the negative shifts resulting from cubic anharmonicity is essential. Building on the renormalized phonon energies obtained from the above self-consistent phonon (SCP) calculations, the further inclusion of cubic anharmonicity can be quantified using the following self-consistent equation within the quasi-particle approximation (*80*):

$$\left(\Omega_q^B\right)^2 = \Omega_q^2 - 2\Omega_q \text{Re} \sum_q^B[G, \Phi_3]\left(\Omega = \Omega_q^B\right), \tag{3}$$

where, $\sum_q^B[G, \Phi_3](\Omega_q)$ denotes the phonon frequency-dependent bubble self-energy, B represents the bubble diagram, G is the phonon propagator and $\Phi_3$ is the third-order force constant, explicitly included in the anharmonic self-energy calculations. Note that the Quasiparticle Nonlinear (QP-NL) treatment was used to solve Equation (3) because of its reliable predictions and the q-mesh used for the self-consistent phonon (SCP) calculations was set to a 2×2×2 configuration.

**Wigner transport formula**

To account for both the population and coherence contributions to the total thermal conductivity, $\kappa_L$, under the single-mode relaxation time approximation (SMRTA), the Wigner transport formula can be utilized, as expressed in the following equation (*53*):

$$\kappa_L^{P/C} = \frac{\hbar^2}{k_B T^2 V N} \sum_q \sum_{j,j'} \frac{\Omega_{qj} + \Omega_{qj'}}{2} v_{qjj'} \otimes v_{qj'j} \cdot \frac{\Omega_{qj} n_{qj}(n_{qj}+1) + \Omega_{qj'} n_{qj'}(n_{qj'}+1)}{4(\Omega_{qj} - \Omega_{qj'})^2 + (\Gamma_{qj} + \Gamma_{qj'})^2} (\Gamma_{qj} + \Gamma_{qj'}), \tag{4}$$

where, the superscripts P and C represent the contributions from populations and coherences, respectively. In this context, $k_B$ denotes the Boltzmann constant, $T$ stands for temperature, $V$ is the unit-cell volume, $\boldsymbol{v}$ is the group velocity matrix, which includes both intra- and inter-branch terms (*81*), and $j$ is the index of phonon branch. When $j = j'$, the formula calculates the populations' contribution $\left(\kappa_L^P\right)$ as obtained from the Peierls-Boltzmann transport equation (PBTE) results. Otherwise, it determines the coherences' contribution $\left(\kappa_L^C\right)$. The total lattice thermal conductivity, $\kappa_L$, is the sum of the populations' contribution ($\kappa_L^P$) and the coherences' contribution ($\kappa_L^C$). For this study, the q-mesh for three-phonon (3ph) and four-phonon (4ph) scattering processes was set to 12×12×12, and a scalebroad parameter of 0.06 was used, ensuring well-converged results for crystalline $Cs_2InNaCl_6$. Thermal transport calculations, accounting for both population and coherence contributions, were conducted using the ShengBTE (82), FourPhonon packages (83), and our proprietary in-house code (84).

**Multiple-phonon and phonon-isotope scattering rates**

Within the SMRTA, the three-phonon (3ph) $\Gamma_q^{3ph}$ and four-phonon (4ph) $\Gamma_q^{4ph}$ scattering rates can be calculated using the following equations (65, 85):

$$\Gamma_q^{3ph} = \sum_{q'q''} \left\{ \frac{1}{2}\left(1 + n_{q'}^0 + n_{q''}^0\right)\zeta_- + \left(n_{q'}^0 - n_{q''}^0\right)\zeta_+ \right\}, \tag{5}$$

$$\Gamma_q^{4ph} = \sum_{q'q''q'''} \left\{ \frac{1}{6}\frac{n_{q'}^0 n_{q''}^0 n_{q'''}^0}{n_q^0}\zeta_{--} + \frac{1}{2}\frac{(1+n_{q'}^0)n_{q''}^0 n_{q'''}^0}{n_q^0}\zeta_{+-} + \frac{1}{2}\frac{(1+n_{q'}^0)(1+n_{q''}^0)n_{q'''}^0}{n_q^0}\zeta_{++} \right\}, \tag{6}$$

where $\zeta_\pm$ and $\zeta_{\pm\pm}$ are defined as follows:

$$\zeta_\pm = \frac{\pi\hbar}{4N}\left|V^{(3)}(q,\pm q',-q'')\right|^2 \Delta_\pm \frac{\delta(\Omega_q \pm \Omega_{q'} - \Omega_{q''})}{\Omega_q \Omega_{q'} \Omega_{q''}}, \tag{7}$$

$$\zeta_{\pm\pm} = \frac{\pi\hbar^2}{8N^2}\left|V^{(4)}(q,\pm q',\pm q'',-q''')\right|^2 \Delta_{\pm\pm} \frac{\delta(\Omega_q \pm \Omega_{q'} \pm \Omega_{q''} - \Omega_{q'''})}{\Omega_q \Omega_{q'} \Omega_{q''} \Omega_{q'''}}, \tag{8}$$

where the phonon mode $q$ is used as a shorthand for a composite index that includes both the wavevector $\boldsymbol{q}$ and phonon branch $j$. The terms $V^{(3)}(q,\pm q',-q'')$ and $V^{(4)}(q,\pm q',\pm q'',-q''')$ represent the reciprocal representations of third- and fourth-order interatomic force constants (IFCs), respectively. Additionally, $\delta(\Omega)$ denotes the delta function that enforces energy conservation in scattering processes, while the Kronecker deltas $\Delta_\pm$ and $\Delta_{\pm\pm}$ correspond to $\Delta_{q\pm q'-q'',Q}$ and $\Delta_{q\pm q'\pm q''-q''',Q}$, respectively, and ensure momentum conservation.

The phonon scattering term resulting from isotope effects can be written as follows (86):

$$\Gamma_q^{\text{isotope}} = \frac{\pi \Omega_q^2}{2N} \sum_{i \in u.c.} g(i) |e_q^*(i) \cdot e_{q'}(i)|^2 \delta(\Omega - \Omega'), \tag{9}$$

where, the mass variance $g(i)$ is defined as $g(i) = \sum_s f_s(i)[1 - M_s(i)/\bar{M}(i)]^2 = \sum_s f_s(i)[\Delta M_s(i)/\bar{M}(i)]^2$, where $f_s(i)$ and $M_s(i)$ represent the concentration and mass of the $sth$ isotope of atom $i$, respectively. $\bar{M}(i)$ denotes the average mass of the $ith$ atom in the primitive cell, and $e_q(i)$ denotes the eigenfunction of phonon mode $q$ at atom $i$.

Using Matthiessen's rule, the total phonon scattering rate $\Gamma_q$ can be expressed as the sum of the individual scattering rates from different mechanisms, given by:

$$\Gamma_q = \Gamma_q^{3\text{ph}} + \Gamma_q^{4\text{ph}} + \Gamma_q^{\text{isotope}} \tag{10}$$

**Acknowledgments**

Y.Z. thanks the Equipment Competition fund (REC20EGR14) and the open fund from the State Key Laboratory of Clean Energy Utilization (ZJUCEU2022009) and the ASPIRE Seed Fund (ASPIRE2022#1) from the ASPIRE League. Y.Z. thanks for the Research Grants Council of the Hong Kong Special Administrative Region under Grant 260206023, C6020-22G and C7002-22Y. Y.Z. thanks for the Hong Kong SciTech Pioneers Award from the Y-LOT Foundation. Y.Z. also thanks the fund from the Guangdong Natural Science Foundation under Grant No. 2024A1515011407. The authors are grateful to the Materials Characterization and Preparation Facility (MCPF) of HKUST for their assistance in experimental characterizations.

**Author contributions**

Y.Z. conceived the idea and supervised the project; G.W. designed the experiments and conducted the material synthesis, characterization, and performance investigation; J.Z. and Y.X. did the calculations; G.W. and J.X. prepared the samples; G.W., J.Z. and Y.Z. prepared the manuscript.; All the authors reviewed and revised the manuscript.

**Supplementary information**

The online version contains available supplementary information.